\documentclass[prd,amsmath,twocolumn]{revtex4-2}
\usepackage{graphicx}
\usepackage{float}
\usepackage{epstopdf,cancel}
\usepackage{epsf,latexsym,bbm,euscript}
\usepackage{amssymb,amsmath,mathtools}
\usepackage{mathtools} 
\usepackage{times,graphics}
\usepackage{soul,xcolor}
\usepackage{mathtools}
\usepackage{mathrsfs}
\usepackage[caption=false]{subfig}
\newcommand{\subfigimg}[3][,]{%
	\setbox1=\hbox{\includegraphics[#1]{#3}}
	\leavevmode\rlap{\usebox1}
	\rlap{\hspace*{-10pt}\raisebox{.5\baselineskip}{\small{#2}}}
	\phantom{\usebox1}
}

\pdfoutput=1
\def\6{{\langle}}
\def\9{{\rangle}}
\newcommand{\defeq}{\vcentcolon=}
\newcommand{\eqdef}{=\vcentcolon}

\newcommand{\be}{\begin{equation}}
\newcommand{\ee}{\end{equation}}
\newcommand{\ba}{\begin{eqnarray}}
\newcommand{\ea}{\end{eqnarray}}

\def\sg{\textsl{g}}

\def\etal{\textit{et al.}}

\def\half{{\tfrac{1}{2}}}

\def\pad{{\partial}}

\def\sg{\textsl{g}}

\def\cO{\mathcal{O}}

\definecolor{orcidlogocol}{HTML}{A6CE39}

\usepackage{scalerel}
\usepackage{tikz}
\usetikzlibrary{svg.path}
\definecolor{orcidlogocol}{HTML}{A6CE39}
\tikzset{
	orcidlogo/.pic={
		\fill[orcidlogocol] svg{M256,128c0,70.7-57.3,128-128,128C57.3,256,0,198.7,0,128C0,57.3,57.3,0,128,0C198.7,0,256,57.3,256,128z};
		\fill[white] svg{M86.3,186.2H70.9V79.1h15.4v48.4V186.2z}
		svg{M108.9,79.1h41.6c39.6,0,57,28.3,57,53.6c0,27.5-21.5,53.6-56.8,53.6h-41.8V79.1z M124.3,172.4h24.5c34.9,0,42.9-26.5,42.9-39.7c0-21.5-13.7-39.7-43.7-39.7h-23.7V172.4z}
		svg{M88.7,56.8c0,5.5-4.5,10.1-10.1,10.1c-5.6,0-10.1-4.6-10.1-10.1c0-5.6,4.5-10.1,10.1-10.1C84.2,46.7,88.7,51.3,88.7,56.8z};
	}
}
\newcommand\orcidlink[1]{\href{https://orcid.org/#1}{\mbox{\scalerel*{
				\begin{tikzpicture}[yscale=-1,transform shape]
					\pic{orcidlogo};
				\end{tikzpicture}
			}{X}}}}

\usepackage{url,hyperref}
\hypersetup{colorlinks,linkcolor={blue!55!black},citecolor={red!50!black},urlcolor={blue!45!black}}

\begin{document}

	\title{Black hole as spherically-symmetric horizon-bound objects}
	
	\author {Pravin K. Dahal\orcidlink{0000-0003-3082-7853}}
	\email{pravin-kumar.dahal@hdr.mq.edu.au}

	\author{Fil Simovic\orcidlink{0000-0003-1736-8779}}
	\email{fil.simovic@mq.edu.au}
	
	\author{Ioannis Soranidis\orcidlink{0000-0002-8652-9874}}
	\email{ioannis.soranidis@hdr.mq.edu.au}

	\author{Daniel R.\ Terno\orcidlink{0000-0002-0779-0100}}
	\email{daniel.terno@mq.edu.au}
	\affiliation{Department of Physics and Astronomy, Macquarie University, Sydney, New South Wales 2109, Australia}
	
	\begin{abstract}
		 Working in a semi-classical setting, we consider solutions of the Einstein equations that exhibit light trapping in finite time according to distant observers. In spherical symmetry, we construct near-horizon quantities from the assumption of regularity of the renormalized expectation value of the energy-momentum tensor, and derive explicit coordinate transformations in the near-horizon region. We examine the boundary conditions appropriate for embedding the model into a cosmological background, describe their evaporation in the linear regime and highlight the observational consequences, while also discussing the implications for the laws of black hole mechanics.

	\end{abstract}
	\maketitle
	
	\section{Introduction}
Dozens of astrophysical black holes (ABHs) --- dark, massive, ultra-compact objects --- exist in the observable Universe. They range in appearance from the components of binary systems with the mass of a few suns, to the supermassive cores of quasars in the centers of galaxies. Beginning with the groundbreaking infrared observations of \cite{GMBTK:00,Sall:02}, ABH candidates are now routinely identified and characterized via gravitational wave interferometry (the LIGO/Virgo collaboration \cite{LIGO:21}), and through electromagnetic observations \cite{B:17}, including very long baseline interferometry in the microwave (the Event Horizon Telescope \cite{EHT:19}), x-ray spectroscopy (using the K$\alpha$ line of iron \cite{IM:19,RN:03}), and more.


As the existence of ABHs is now established beyond reasonable doubt,  the question of their physical nature \cite{BCNS:19,CP:19,M:23} becomes relevant.
Broadly speaking, there are two competing views on the nature of ABHs. The first view identifies them as  mathematical black holes (MBHs). Their defining  feature is the event horizon, a null surface that causally disconnects the black hole interior from the outside world. For    the Schwarzschild black hole solution it is located at the gravitational radius $r_\sg=2GM/c^2$.  MBHs are possibly the most dramatic prediction of general relativity  and embody our traditional notions of black holes \cite{exact:03,MTW,HE:73,FN:98,vF:15,HMTC:16}.  The MBH paradigm explains a staggering variety of astrophysical phenomena and successfully models ABH properties across all currently accessible time and length scales \cite{BCNS:19,CP:19,FCZT:23}.

Nevertheless, identifying ABHs with MBHs comes with a conceptual price.  The exteriors of Schwarzschild or Kerr MBHs are regular, but their interiors are not. They contain Cauchy horizons and singularities, such as the curvature singularity of the Schwarzschild solution at $r=0$. Such pathologies are expected to be resolved by a presently unknown quantum theory of gravity, but   the known quantum effects are responsible for a host of technical difficulties and unresolved paradoxes \cite{FN:98,M:15,H:16,BMPS:95,info:21}.

All of the above motivates the second view which postulates  existence of some black hole mimickers that fit the observed data (and are thus sufficiently close to the MBH solutions of general relativity), but are pathology-free. A variety of models \cite{BCNS:19,CP:19,CP:17} designated as horizonless exotic compact objects (ECOs) appear to provide an  alternative explanation of the observed ABHs, at the price of modifying known physics and/or the introduction of some exotic quantum matter.

This conceptual dichotomy is somewhat blurred \cite{CDLV:23}, especially if we take into account the following. On the one hand,  Schwarzschild or Kerr geometries are the asymptotic states of a classical gravitational collapse. According to a distant observer (who we refer to as Bob) once  the stellar remnant cannot be supported by degeneracy pressure,  it turns into  a frozen dark star of radius $r\approx r_\sg$ within a few light crossing times $t\sim r_\sg/c$. However,  the event horizon is in principle an unobservable teleological entity \cite{H:00,mV:14,J:14}, and quantum effects may prevent it from forming at all \cite{H:14,C-R:18}.  Both numerical and observational studies thus focus on other characteristics of black holes \cite{HMTC:16,RZ:13}. This is the rationale behind ECOs --- they are designed to closely mimic a MBH  without forming an event horizon \cite{CP:19}.

However, the conceptual price of this mimicking is the violation of one or more of the natural assumptions entering Buchdahl's theorem \cite{CP:19,B:59}. A direct or indirect result of these violations is the existence of non-classical matter, whose energy-momentum  tensor (EMT) $T_{\mu\nu}$ violates at least the null energy condition (NEC), which states  $T_{\mu\nu}k^\mu k^\nu\geqslant 0$ for all null vectors $k$ \cite{HE:73,vF:15,KS:20}. On the other hand, the existence of Hawking radiation   leads to a large but finite lifetime for black holes and itself violates the NEC in the vicinity of the apparent horizon \cite{FN:98}. This motivates the introduction of another class of singularity-free objects, regular black holes (RBHs), which represent domains of spacetime that enable temporary but prolonged trapping of light \cite{H:06,F:16,CDLPV:18,M:22}. The trapping of light underpins our notions of what physically constitutes a black hole \cite{eC:19} and we use it as its defining feature \cite{F:14,MMT:22}.

It is useful to introduce a suitable (not uniquely defined) parameter $\epsilon$ that characterizes how close a proposed ultra-compact object is  to its Schwarzschild or Kerr MBH with the same mass and spin \cite{CP:17}.  The behaviour of $\epsilon$ allows one to select among three different types of models: (i) Classical MBHs correspond to the asymptotic scenario in which $\epsilon\to 0$ as $t\to \infty$, where $t$ is the time measured by Bob. (ii) Various ECOs correspond to an $\epsilon>0$ that is reached at finite $t$ or approached asymptotically. (iii) Finally, an evaporating RBH is a particular example of a trapped spacetime region (with $\epsilon=0$) that forms in a finite time according to Bob, i.e. $\epsilon=0$ for some $t<\infty$. The definition of a {\it physical} black hole (PBH) as a trapped spacetime domain was introduced in Ref.~\cite{F:14}.  We supplement this definition by the additional operationally motivated requirement that the trapped region form in finite time according to Bob \cite{MMT:22}.  Such a PBH may or may not have an event horizon or singularity \cite{MMT:22,DST:22}.

As a result there is a need to distinguish between three classes of models that describe astrophysical black holes --- MBHs, ECOs, and PBHs \cite{M:23,MMT:22} --- each with their own defining features.
The   event horizon is the most recognizable conceptual characteristic of a (mathematical) black hole.  In  numerical relativity black holes are identified via apparent horizons (part of the definition of a physical black hole). Exotic compact objects are models that dispense with the horizon altogether.
 To uncover the true  nature of ABHs it is therefore necessary to compare the  properties of PBHs with those of conventional   semiclassical black holes, and identify the potential for extracting observational signatures.

This comparison,  however, cannot  take place in  an asymptotically flat spacetime to which the standard MBH solutions belong. Current observations indicate that within sub-percent precision, the Universe is described at cosmological scales by the
perturbed spatially flat Freidmann--Robertson--Lema\^{\i}tre--Walker (FRLW) metric \cite{vM:05,planck6}. The Kerr solution is asymptotically flat and is thus necessarily provisional, even if the issues surrounding singularities and event horizons are resolved. Beyond time and length scales that are small relative to the reciprocal Hubble parameter $H$, it can only be treated as an approximation to a more general solution \cite{FCZT:23}. In a separate but related development, activity over the last two decades has led to a renewed interest in mathematical models of inhomogeneities in the cosmological background, which straddle the cosmological and black hole scales \cite{FGF:21}.

This work represents the first in a three part series of papers aimed at addressing these issues. In this first part, we take steps towards developing a complete framework for modelling astrophysical black holes as PBHs, i.e. objects with horizons that have already formed according to distant observers.  Building on the previous work we complete the description of the near-horizon geometry of a spherically-symmetric PBH.   We demonstrate a general procedure for describing a PBH  as inhomogeneities in the FRLW background, and provide details of their embedding in a spatially flat asymptotically de Sitter spacetime. Since a majority of the results on cosmological black holes \cite{vF:15,FGF:21} and concrete results on PBHs \cite{MMT:22} are obtained in spherical symmetry, we work in this simplifying setting.

To this end we review the main aspects of the formalism used to construct the PBH model in Sec.~\ref{pbh}. One important feature of a spherically-symmetric PBH is that its growth is impossible, and only contraction (usually referred as evaporation) is allowed.  In Sec.~\ref{pbhB} we derive general relations between the leading contributions to near-horizon quantities in the two systems best adapted to evaporating BH models --- $(t,r)$ and $(v,r)$ coordinates. In Sec.~\ref{model}, we present exact solutions for the case of linear evaporation, and show that a linear evaporation law in one coordinate system necessarily implies linear evaporation in the other. In Sec.~\ref{cosmo}, we show that the PBH metric can be consistently embedded in an FRWL cosmology, and propose a representative compactification of the resulting spacetime.  We conclude in Sec.~\ref{discus} with a summary of our results, their implications, and directions for future work. Throughout, we work in units where $\hbar=c=G=1$.

		\section{Spherically-symmetric physical black holes}\label{pbh}
		
\subsection{General set-up and admissible solutions}

The self-consistent approach \cite{MMT:22} is based on semiclassical gravity \cite{HV:20}. The spacetime geometry is described by a metric $\sg_{\mu\nu}$, and the notion of test particles' trajectories, horizons, etc. are assumed to be well-defined. The metric itself is a solution of the Einstein equations, which may include higher-order curvature terms and a cosmological constant. Their source is the energy-momentum tensor $T_{\mu\nu}\defeq\6\hat T_{\mu\nu}\9_\omega$, which is a renormalized expectation value of some EMT operator in some unspecified state of gravity and matter $\omega$. We do not make any assumption about the nature of matter fields or their quantum states, and do not separate the background (cosmological and/or collapsing matter) from the generated quantum excitations. The goal is to infer as much information as possible about the EMT and the metric in the vicinity of the apparent horizon simply from its existence.

Thus in practice we analyse the behaviour of solutions to
\be
R_{\mu\nu}-\half \sg_{\mu\nu}R=8\pi T_{\mu\nu}\ , \label{EE}
\ee
where $R_{\mu\nu}$ and $R$ are the Ricci tensor and scalar, respectively, and the right hand includes some or all of the described above components.

A general spherically symmetric metric in Schwarzschild coordinates \cite{MTW,vF:15} is given by
\be
ds^2 = -e^{2h(t,r)}f(t,r)dt^2+f(t,r)^{-1}dr^2+r^2d\Omega_2\ , \label{m:tr}
\ee
while using the advanced null coordinate $v$ results in the form
\be
ds^2=-e^{2h_+(v,r)}f_+(v,r)dv^2+2e^{h_+(v,r)}dvdr+r^2d\Omega_2\ . \label{m:vr}
\ee
The function $f$ is coordinate-independent, i.e. $f(t,r)\equiv f_+\big(v(t,r),r\big)$ and in what follows we omit the subscript. It is conveniently represented via the Misner--Sharp--Hernandez (MSH) mass $M\equiv C/2$ as
\be
f=1-\frac{C(t,r)}{r}=1-\frac{C_+(v,r)}{r}=\pad_\mu r \pad^\mu r\ ,
\ee
where the coordinate $r$ is the areal radius \cite{vF:15}. 		
The functions $h$ and $h_+$ play the role of integrating factors in the coordinate transformation
\be
dt=e^{-h}(e^{h_+}dv-f^{-1}dr)\ . \label{trvr-transformation}
\ee
In an asymptotically flat spacetime, $h\to 0$ and $f\to 1$ as $r\to 0$, and $t$ is the physical time of a stationary observer Bob at spacelike infinity $i^0$. For example, the Schwarzschild metric corresponds to $h\equiv 0$, $M\equiv C/2=\mathrm{const.}$, and $v=t+r_*$, where $r_*$ is the tortoise coordinate \cite{MTW,HE:73}.  A description in terms of the retarded null coordinate $u=t-r_*$ and its properties are described in Appendix \ref{ur-section}.

A PBH is a trapped region --- a  domain where both
ingoing and outgoing future-directed null geodesics emanating
from a spacelike two-dimensional surface with
spherical topology have negative expansion \cite{HE:73,vF:15,S:11}. The apparent horizon is the
boundary of this trapped region. In a cosmological setting, we assume that a separation of scales exists between geometric features associated with the black hole and those of the large-scale universe. In this case, the apparent horizon is given by the outermost real root of $f(t,r)=0$ in the near-region, while the cosmological horizon is the innermost real root in the asymptotic region (the detailed summary of various definitions can be found in Refs.~\cite{vF:15,MMT:22,S:11}.
In an asymptotically flat spacetime the Schwarzschild radius $r_\sg$ is the largest root of $f(t,r)=0$.  Invariance of the MSH mass implies that
\be\label{msh}
r_\sg(t)=C(t,r_\sg)=r_+\big(v(t,r_\sg(t)\big)\big),
\ee
where $r_+(v)$ is the largest root of $f_+(v,r)=0$. It represents the location of the outer component of the apparent horizon. Unlike the globally defined event horizon, the notion of the apparent horizon is foliation-dependent. However, it is invariantly defined in all foliations that respect spherical symmetry \cite{FEFHM:17}.

In addition to requiring that a PBH is formed in a finite time according to Bob, we demand only the weakest form of the cosmic  censorship conjecture \cite{FN:98,vF:15,P:79}: all curvature scalars \cite{HE:73,exact:03} are finite up to and on the apparent horizon. It sufficient to ensure that only two of them, $R$ and $R_{\mu\nu}R^{\mu\nu}$, are finite \cite{T:19}. Construction of finite invariants from the divergent quantities that describe a real-valued solution allows one to describe properties of the near-horizon geometry. Because the metric in Schwarzschild coordinates is singular at the apparent horizon, it will often be convenient to work in null coordinates instead.

Both the analysis of the Einstein equations and the evaluation of curvature invariants is  conveniently performed using the   effective EMT components $\tau_a$, (where $a={}_t, {}^r, {}_t{}^r $)  defined as \cite{MMT:22}
\begin{align}
	\tau{_t} \defeq e^{-2h} {T}_{tt}\ , \qquad {\tau}{^r} \defeq T^{rr}\ , \qquad  \tau {_t^r} \defeq e^{-h}  {T}{_t^r}\ . \label{eq:mtgEMTdecomp}
\end{align}
The Einstein equations for the components $G_{tt}$, ${G}{_t^r}$, and ${G}^{rr}$ are then, respectively
\begin{align}
	\partial_r C &= 8 \pi r^2  {\tau}{_t} / f\ , \label{eq:Gtt} \\
	\partial_t C &= 8 \pi r^2 e^h  {\tau}{_t^r}\ , \label{eq:Gtr} \\
	\partial_r h &= 4 \pi r \left(  {\tau}{_t} +  {\tau}{^r} \right) / f^2\ . \label{eq:Grr}
\end{align}
To ensure finite values of the curvature scalars, it is sufficient to work with only two invariant quantities
\be
\bar{\mathrm{T} }\defeq\mathrm{T} +2T^\theta_{~\theta}, \qquad \bar{\mathfrak{T}}\defeq\mathfrak{T}+2\big(T^\theta_{~\theta}\big)^2,
\ee
where
\begin{align}
	\mathrm{T} &\defeq T^\mu_\mu= ( {\tau}{^r} -  {\tau}{_t}) / f\ ,
\end{align}
\begin{align}
	\mathfrak{T} \defeq T^{\mu\nu}T_{\mu\nu} =\big( ( {\tau}{^r}&)^2 + ( {\tau}{_t})^2 - 2 ( {\tau}{_t^r})^2 \big) / f^2\ ,
	\label{eq:TwoScalars}
\end{align}
where the contributions of $T^\theta_{~\theta}\equiv T^\phi_{~\phi}$ are disregarded, as one can verify that they do not introduce further divergences \cite{MMT:22,T:19}.

These considerations restrict the scaling of the effective EMT components near the apparent horizon, such that $\tau_a\propto f^k$, with $k=0,1$. Solutions with $k=0$ describe a PBH after  formation (and before a possible disappearance of the trapped region).  Dynamical RBH solutions belong to this class \cite{MS:23}, while the Reissner-Nordstr\"{o}m solution or static RBH solutions correspond to $k=1$.  In the following we will almost exclusively work with $k=0$ solutions.

The admissible (i.e. real-valued) $k=0$ solutions satisfy
\be
\lim_{r\to r_\sg} \tau_t=\lim_{r\to r_\sg} \tau^r=-\Upsilon^2(t)\ , \quad \lim_{r\to r_\sg} \tau^r_t=\pm \Upsilon^2(t)\ ,
\ee
for some function $\Upsilon(t)$.
The leading terms of the metric functions  are given in terms of $x\defeq r-r_\sg(t)$ as
\begin{align}
	C &= r_\sg - 4 \sqrt{\pi} r_\sg^{3/2} \Upsilon \sqrt{x} + \mathcal{O}(x)\ , \label{k0C} \\
	h &= - \frac{1}{2}\ln{\frac{x}{\xi}} + \mathcal{O}(\sqrt{x})\ . \label{k0h}
\end{align}
The function $\Upsilon(t)$ determines the energy density, pressure and flux at the apparent horizon, and  $\xi(t)$ is determined by choice of the time variable. The higher-order terms are matched with higher-order terms in the EMT expansion \cite{BsMT:19,MMT:22}.

The Einstein equation~\eqref{eq:Gtr} serves as a consistency condition and establishes the relation between the rate of change of the MSH mass and the leading terms of the metric functions,
\begin{align}
	r'_\sg/\sqrt{\xi} =   \pm 4\sqrt{\pi r_\sg} \, \Upsilon\ , \label{eq:k0rp}
\end{align}
where primes indicate derivatives with respect to $t$ and $ \pm $ sign corresponds to the expansion and contraction of the Schwarzschild sphere, respectively.
For a  contracting Schwarzschild sphere   the $(v,r)$ coordinates are regular across it. Evaluation of the expansion of the geodesic congruences identifies the solutions with $r_\sg '<0$ as  black holes of decreasing mass. Similarly,  the case $r_\sg'>0$  allows for a regular description in $(u,r)$ coordinates. Then the region $f(u,r)<0$ is anti-trapped, and the solution describes an expanding white hole. In the following we consider only PBHs.

\begin{figure*}[!htbp]
	\centering
	\vspace{-30mm}
	\begin{tabular}{@{\hspace*{+0.1\linewidth}}p{0.05\linewidth}@{\hspace*{0.35\linewidth}}p{0.45\linewidth}@{}}
		\centering
		\subfigimg[scale=0.6]{(a)}{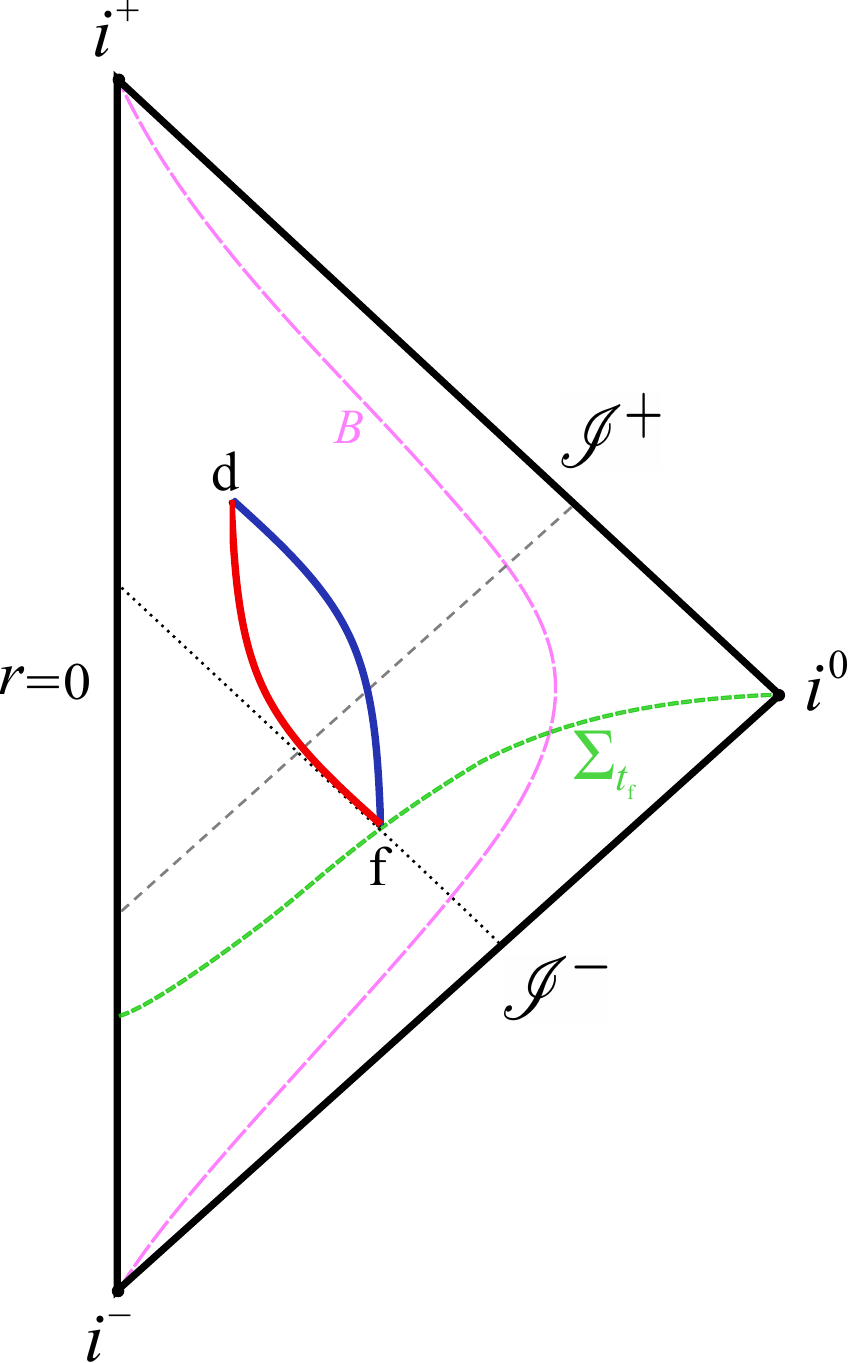} &
		\subfigimg[scale=0.6]{(b)}{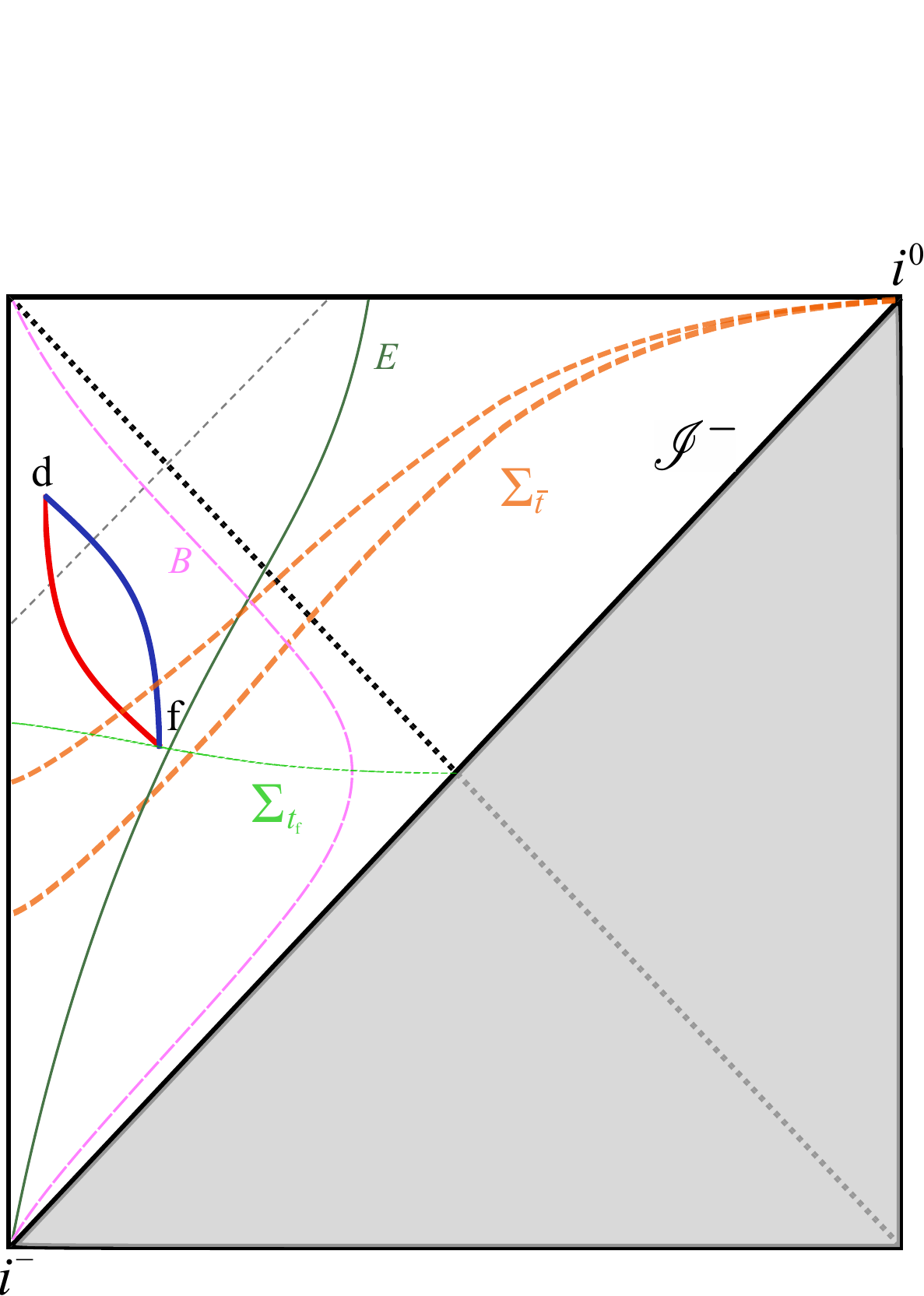}
	\end{tabular}
	\vspace{0mm}
	\caption{Schematic Carter–Penrose diagram depicting the formation and evaporation of a RBH which is treated as a particular case of a PBH.  Past and future timelike infinity are labelled by $i^-$ and $i^+$, respectively. Spacelike infinity is labelled by $i^0$.
		Dashed grey lines correspond to outgoing radial null geodesics. The trajectory of a distant observer, Bob, is indicated in pink and labelled $B$. The points $\mathrm{f}$ and $\mathrm{d}$ represent the two-spheres of formation and disappearance of the trapped region. The equal (Schwarzschild) time hypersurface $\Sigma_{t_{\mathrm{f}}}$ is shown as a dashed light green line.  The outer (blue) and inner (dark red) components of the black hole apparent horizon (timelike membranes) are indicated according to the invariant definition (\cite{vF:15,MMT:22}. (a) The invariantly-defined components of the apparent horizon correspond   the largest and smallest root of $f=0$ whether $t$, $v$ or $u$ is used as the
		evolution parameter \cite{DST:22}.   (b) Embedding into de Sitter spacetime. The solid black line connecting $i^{-}$ and $i^{0}$ represents the cosmological event horizon for an observer at $r=0$. Static coordinates cover only the left quadrant, with the dotted diagonal line representing the particle horizon. Components of the black hole apparent horizon correspond to the largest and smallest roots of $f=0$ (not including the cosmological horizon). The orange dashed lines $\Sigma_{\bar{t}}$ indicate hypersurfaces of constant comoving time $\bar t$. The trajectory of an asymptotically comoving observer Eve ($\chi=\mathrm{const}$)  is marked by the dark green line and labelled by the initial $E$.  }
	\label{rbh2}
\end{figure*}

PBH metrics in Schwarzschild coordinates are more singular than the  Schwarzschild or   Reissner-Nordstr\"{o}m solutions. Unlike the special algebraic case \cite{exact:03,FGF:21} $\sg_{tt}=\sg_{rr}=1$, the metric determinant $\sg\defeq\det\sg_{\mu\nu}$ diverges as $x^{-1}$ on  approach to the apparent horizon. The   EMT near the Schwarzschild sphere is
\begin{align}
	\ T^a_{~b} = \begin{pmatrix}
		\Upsilon^2/f &   e^{-h}\Upsilon^2/f^2 \vspace{1mm} & 0 & 0\\
		- e^h  \Upsilon^2 & -\Upsilon^2/f & 0 &0 \\
 0 & 0 & p_\| &0 \\
 0 & 0 & 0 & p_\|
	\end{pmatrix},
		\label{tneg}
\end{align}
where the tangential pressure $p_\|$ is finite at $r=r_\sg$, and for a static observer $p=\rho= -\Upsilon^2/f+\cO(f^0)$ as $r\rightarrow r_\sg$.  Writing the $(t,r)$ block in an orthonormal frame \cite{MMT:22},
\begin{align}
	T_{\hat{a}\hat{b}} = -\frac{\Upsilon^2}{f} \begin{pmatrix}
		1 &1 \vspace{1mm}\\
		1   & 1
	\end{pmatrix},
\end{align}
makes violation of the null energy condition particularly transparent.

 For a static $r=\mathrm{const.}$ observer that we call Eve, the energy density, pressure, and flux all diverge. Direct transformations show that in $(v,r)$ coordinates all of the EMT components are finite \cite{T:19,MMT:22}. In particular, if we choose the advanced null coordinate in such a way that $\zeta_0=0$ (see Eq.~\eqref{h1} below), then $T_{vv}|_{r_+}=-\Upsilon^2$. For the Vaidya metric other EMT components are zero, resulting in additional relations between the higher-order terms \cite{DST:22} in $(t,r)$ coordinates. We note that this self-consistent approach by definition constrains the expectation value of the total EMT, but by itself does not describe either the collapsing matter content nor the spectral representation of the resulting quantum excitations. 

The Schwarzschild sphere $r_\sg(t)$ is a timelike hypersurface \cite{T:19,MMT:22}. Therefore, ingoing null geodesics and some of the ingoing timelike geodesics can cross the apparent horizon in a finite time according to Bob. Indeed,  ingoing radial null geodesics satisfy
\be
\left.\frac{dr}{dt}\right|_{v=\mathrm{const}}= -e^hf\ ,
\ee
so by noting that
\be
\lim_{r\to r_\sg}e^h f=|r_\sg'| \ ,
\ee
(instead of diverging as $1/f$ in the case of the Schwarzschild black hole),  we see that the infall into a PBH takes a finite (even if very large) time according to Bob \cite{MMT:22,T:19,BsMT:19}.

In $(v,r)$ coordinates the black hole metric is described by
\begin{align}\label{eq:Cpl}
	C_+(v,r) &= r_+(v)+w_1(v)y+\cO(y^2)\ ,\\
	h_+(v,r) &= \zeta_0(v)+\zeta_1(v) y+\cO(y^2)\label{h1} \ ,
\end{align}
where $y\defeq r-r_+(v)$.
Note that a freedom in the redefinition of the null variable $v$ allows one to set $\zeta_0\equiv 0$. 		
From the definition of the apparent horizon it follows that $w_1\leqslant 1$. The inequality is saturated at the formation of the PBH  (more details can be found in \cite{DST:22}). Note that in the $(v,r)$ coordinates the Schwarzschild solution $h_+\equiv 0, C_+(v,r)\equiv r_+$ is a limiting case of dynamical metrics. On the other hand in $(u,r)$ coordinates,  the metric function $h_-(u,r)$ diverges at the apparent horizon of $k=0$ solutions  (see Appendix \ref{ur-section})

It is easy to see that the Schwarzschild sphere at $r=r_\sg(t)=r_+(v)$ is timelike. Similarly, if the equation $f(t,r)=0$ has more than one solution for $r\ll 1/H$, the innermost surface $r_{\text{in}}$ is timelike as well. As a result, these definitions of the inner and outer horizons coincide with the invariant definitions \cite{DST:22}, and a RBH in an asymptotically flat spacetime has the schematic Carter--Penrose diagram shown in \figurename{ \ref{rbh2}(a)}.

In spherical symmetry the black hole mass is defined as the value of the MSH mass at the outer apparent horizon \cite{vF:15},
\be
2M(t)\defeq C\big(t,r_\sg(t)\big)\equiv r_\sg(t) \ ,
\ee
with analogous expressions holding in $(v,r)$ and $(u,r)$ coordinates. This is consistent with  Eq.~\eqref{k0C} that also leads  to    
\be
\frac{dC(t,r_\sg)}{dt}=r'_\sg(t)\ .
\ee

Finally we define the notion of surface gravity, which plays a crucial role in black hole mechanics and thermodynamics \cite{HE:73,FN:98,vF:15}. For stationary black holes the various definitions that appear in the literature are equivalent, but this degeneracy is lifted in the dynamical case. Among the possible generalizations, the Hayward--Kodama surface gravity \cite{vF:15,VAD:11}
\be
\kappa_{\mathrm{K}}
=\frac{1}{2}\left(\frac{C_{+}(v,r)}{r^2}-\frac{\partial_r C_{+}(v,r)}{r}\right)\!\bigg|_{r_+}\!\!\! =\frac{(1-w_1)}{2r_+}\label{kodama}
\ee
stands out as the most useful candidate \cite{MMT:22D}. It is based on the Kodama vector \cite{K:80}, which provides a preferred time flow in the absence of a timelike Killing vector field.  It also allows for the generalization of the first law of black hole mechanics to dynamical spacetimes \cite{Hayward:98}, since the Kodama vector is associated with a conserved current. In fact its Noether charge is just the MSH mass defined previously, which in black hole thermodynamics plays the role of the internal energy of the system. Unlike some alternatives, it is well-defined for the PBH and shares many of the important properties of its stationary Killing counterpart, and will be used throughout this work.

		\subsection{Some properties of physical black holes}\label{pbhB}

		PBHs can be described in both $(t,r)$ and $(v,r)$ coordinates, as seen from Eqs.~\eqref{m:tr} and \eqref{m:vr}. In this section we examine  connections between the metric functions in these coordinates   using the transformation law \eqref{trvr-transformation}. In $(t,r)$ coordinates the MSH mass is given by the expansion \eqref{k0C}, while in $(v,r)$ coordinates it is given by \eqref{eq:Cpl}. We examine the relation (close to the apparent horizon) connecting the quantities $x$ and $y$ and determine what information can be extracted from the invariance of the MSH mass.
		
		While the metric in $(t,r)$ coordinates is singular at the apparent horizon, a freely-falling observer Alice reaches the apparent horizon at $r_\sg$ not only in her finite proper time $\tau$ but in finite $t$.  We thus can consider the change in  $t$ from the value $t\big(v,r_+(v)\big)$ along an ingoing null geodesic $v=\mathrm{const}$.  Along such a geodesic the time $t(v,r)$ varies as
		\be\begin{aligned}
			t(v,r_++y)=t(v,r_{+})+\left.\partial_{r}t\right|_{r_+}y+\tfrac{1}{2}\left.\partial^2_{r}t\right|_{r_+}y^2+\cO(y^3)\ .
		\end{aligned}\ee
		Determining the explicit form of the above relation requires evaluating partial derivatives at the apparent horizon. This can be done using the transformation law \eqref{trvr-transformation}, which implies directly that
		\begin{align}
			\partial_{r}t=-e^{-h(t,r)}f(t,r)^{-1}=\frac{1}{r'_{g}}+\cO(\sqrt{x})\ .
		\end{align}
		 The time variation  $\delta t\defeq	t(v,r_{+}+y)-t(v,r_{+})\,$ along an ingoing null geodesic is thus given by
		\begin{align}
			\delta t=\frac{y}{r'_{g}}+\tfrac{1}{2}(\partial^2_{r}t)\!\!\!\underset{y=0}{\big|}y^2+\cO(y^3)\label{delta-t}\ ,
		\end{align}
where the second partial derivative $(\partial^2_{r}t)$ is given in Appendix \ref{appendix-d2tdr2}. 	The corresponding expansion of the  Schwarzschild radius $r_{\sg}(t)$  is given by
		\be\begin{aligned}
			r_{\sg}(t(v,r_{+}+y))&=r_{\sg}(t(v,r_{+}))+r'_{g}\delta t\\
			&\quad+\tfrac{1}{2}r''_{\sg}\delta t^2+\cO(\delta t^3)\ , \label{rg-var}
		\end{aligned}\ee
	  where keeping terms of order $\delta t^2$ is crucial.

		The variable $x(t,r)=r-r_\sg(t)$ can further be expressed as a function of the advanced null coordinate $v$ and $r$,
		\be
 x(v,r_{+}+y)=(r_{+}+y)-r_{\sg}(t(v,r_{+}+y))\ , \label{x-gen}
		\ee
Using Eqs.~\eqref{delta-t} and \eqref{rg-var} in \eqref{x-gen} along with the invariance of the MSH mass \eqref{msh} then results in the quadratic relationship between $x$ and $y$ near the apparent horizon:
	 \begin{align}\label{xy}
	 	x=\tfrac{1}{2}\, \omega^2y^2 \ ,\quad\text{where}\quad  \omega^2\equiv-r'_{g}(\partial^2_{r}t)\!\!\!\underset{y=0}{\big|}\!\!-\frac{r''_{g}}{(r'_{g})^2}
	 \end{align}
Then by using Eqs.~\eqref{rg-var} and \eqref{xy} along with \eqref{msh} 
     we find that
    \begin{align}
    	w_{1}(v)=1- 2\sqrt{2\pi \smash{r_{g}^{3}}\vphantom{r_{g}^{3}}}\,\Upsilon\, {\omega}\ ,\label{w1v}
    \end{align}
    which is the quantity entering the Hayward--Kodama surface gravity in Eq.~\eqref{kodama}.
	 Explicit expressions  for $\omega^2$ and $w_{1}(v)$ can be found in Appendix \ref{appendix-d2tdr2} and \ref{appendix-vr}.
	
	 We next turn to the evaluation of the unknown metric functions $\Upsilon(t)$ and $\xi(t)$. We assume the evaporation law in $(t,r)$ and $(v,r)$ coordinates can be written as
	 \begin{align}
	 	r'_{g}(t)=-\Gamma(r_{g})\ , \quad r'_{+}(v)=-\Gamma_{+}(r_{+})\,
	 \end{align}
	 in terms of the undetermined functions $\Gamma$ and $\Gamma_+$. The relation \eqref{rpp}, which is derived from the Einstein equations in $(v,r)$ coordinates, determines $\Upsilon(t)$:
	 \begin{align}
	 	\Upsilon(t)=\sqrt{\frac{\Gamma_{+}(1-w_{1})}{8\pi \smash{r^2_{+}}}}\ \label{ups-gen}
	 \end{align}
	 Using the consistency condition Eq.~\eqref{eq:k0rp} along with \eqref{ups-gen} determines the other unknown metric function:
	 \begin{align}
	 	\xi(t)=\frac{r_{g}\Gamma^2}{2\Gamma_{+}(1-w_{1})}
	 \end{align}
	 We now make the following assertion: in the quasi-static limit the first law of black hole dynamics should approach that of the stationary case, where
	 \be\label{firstlaw}
	 dM=\dfrac{\kappa}{8\pi}dA\ ,
	 \ee
	 and the inclusion of electric charge or angular momentum manifests by the appearance of work terms like $\Phi\, dQ$ or $\Omega\, dJ$ which arise in the Hamiltonian variation. The first law of black hole mechanics \eqref{firstlaw} plays a fundamental role in connecting classical geometry with quantum gravitational degrees of freedom, and is known to hold in any diffeomorphism-invariant Lagrangian theory of gravity \cite{bardeen, HE:73,FN:98,IW:94}. If the apparent horizon is taken to be the relevant surface for which the first law is formulated, then the area being $A=4\pi r_\sg^2$ and the MSH mass being $M=r_\sg/2$ requires that $w_1=0$ for the identification of $\kappa$ with the Hayward--Kodama surface gravity of Eq.~\eqref{kodama}. As shown in \cite{hayward1998}, the fully dynamical version of the first law contains a work term with contributions from the trace of the energy-momentum tensor and variation of the apparent horizon volume, but this does not alter the constraint imposed on $w_1$ from demanding the surface gravity to have a static limit of $\kappa \rightarrow 1/2r_\sg$. The physical implications of this constraint for RBHs have been analyzed in \cite{MS:23}. In the case where $w_1=0$, $\Upsilon(t)$ and $\xi(t)$ then assume the forms
     \begin{align}
     	\Upsilon(t)=\sqrt{\frac{\Gamma_{+}}{8\pi \smash{r^2_{+}}}}\ , \quad \xi(t)=\frac{r_{g}\Gamma^2}{2\Gamma_{+}}\ ,\label{Ups-Vaidya}
     \end{align}
	 and $ {\omega}$ as defined in \eqref{xy} reduces to
	 \begin{align}
	 	 {\omega}=\frac{1}{2\sqrt{2\pi \smash{r^3_{g}}\mathstrut}\Upsilon}\ .\label{varpi-xy-w1=0}
	 \end{align}
	 This leads to the following relationship between the near-horizon expansion parameters $x$ and $y$:
	 \begin{align}
	 	x=\frac{1}{16\pi r^3_{g}\Upsilon^2}\,y^2
	 \end{align}

	 The results above follow from the connection between $(t,r)$ and $(v,r)$ coordinates along a constant-$v$ line. Analogous relations can be derived by instead considering the relationship between $(t,r)$ and $(u,r)$ coordinates, as detailed in Appendix \ref{ur-section}. However in this case the retarded null coordinate $u$ exhibits singular behaviour on the apparent horizon in concert with the Schwarzschild coordinate $t$, and the metric functions exhibiting similar behaviour.

We now demonstrate that these results imply that   the near-horizon metric  of the PBH is described by the ingoing Vaidya metric. Having assumed that $w_1=0$, the expansion of the metric functions \eqref{eq:Cpl} and \eqref{h1} become
			\begin{align}
				C_{+}(v,r)&=r_{+}+\cO(y^2)\ ,\\
				h_{+}(v,r)&=\zeta_1(v)y+\cO(y^2)\ ,
			\end{align}
		 where $\zeta_0$ is set to zero by a suitable time reparametrization. Following the semiclassical arguments \cite{jB:81,BMPS:95} that in the quasi-stationary region $\pad h_+/\pad r\sim L_H/r$ (where $L_H$ is the Hawking luminosity) we have that $\zeta_1 r_+\ll 1$, and thus a PBH near the Schwarzschild sphere is well-described by   a Vaidya metric,
		 \begin{align}
		 	ds^2=-f(v,r)dv^2+2dvdr+r^2d\Omega_2\ \label{Vaidya} ,
		 \end{align}
		where
		\begin{align}
			f(v,r)=1-\frac{r_{+}(v)}{r}\ .
		\end{align}

 A common assumption used in various models \cite{APT:19,MMT:22} is that the mass-loss rate in the long-lasting quasi-stationary regime follows Page's law, that
		\be
		\dfrac{dr_\sg}{dt}= -\dfrac{A}{r_\sg^2}\ , \label{Page-2}
		\ee
		for some constant $A$ \cite{page76,FN:98}. This has the same form in $(t,r)$, $(v,r)$ and $(u,r)$ coordinates. Identification of the Hawking temperature with the Kodama--Hayward surface gravity $\kappa_\mathrm{K}$ leads to $w_1=0$. The above analysis is consistent with these assumptions. As a result we identify
\be
\Upsilon^2=\frac{A}{8\pi r_\sg^4}\ , \qquad\xi=\frac{A}{2r_\sg}\ .
\ee

 For macroscopic black holes the evaporation law Eq.~\eqref{Page-2} can be treated as linear for times that are long compared to the cosmological timescale but are still short relative to the evaporation time.  Moreover, the linear Vaidya metric has been proposed as the correct description near the endpoint of the evaporation process \cite{martin2013} and serves as a basis for model-building in the semiclassical setting. This approximation allows for the explicit expressions for coordinate transformations that we now describe.
	
		\subsection{Linear mass loss in Vaidya metric: coordinate transformations}\label{model}
		
Different aspects of the black hole geometry are best captured by different coordinate systems. However, transformations between them are difficult \cite{APT:19}, and where exact coordinate transformations do exist, multiple coordinate patches are required to cover the entire spacetime \cite{ber17}.

A linear dependence of the MS mass on the null coordinate $v$ or $u$ allows for an analytic solution to a number of problems \cite{B:16,WL:86,DST:22}. Here we complement these works by providing an explicit analytic expression for the coordinate transformation from $(v,r)$ to $(t,r)$ coordinates.

For a slowly contracting horizon we have that $r_+'\ll 1$, which holds for the emission of Hawking radiation by a macroscopic black hole and is a good approximation for a sufficiently long interval of $v$. In what follows we assume a linear evaporation law, such that
		\begin{align}
			r_{+}(v)=r_{0}-\alpha v\ , \quad \text{with}\quad\alpha >0\ ,
		\end{align}
		where $r_0$ is the initial areal radius and $\alpha$ is the evaporation rate. The timelike nature of the Schwarzschild sphere allows for arbitrary values of $\alpha$, but the previous considerations restrict  it to $\alpha\ll 1$. While the extension of this metric to large distances $r\gg r_+$ is not justified, it provides a setting in which the exact transformations to $(t,r)$ and $(u,r)$ coordinates are possible. Moreover, its counterpart with decreasing $r_-(u)$ provides a good description of an evaporating black hole at distances $r\gtrsim 2r_\sg$ and its transformation to $(t,r)$ coordinates can be performed analogously.

 The first step in the transformation to Schwarzschild coordinates is to bring the metric into a form that is conformally equivalent to the Schwarzschild metric in $(u,r)$ coordinates. This is effected by defining
		\begin{align}
			v\eqdef\frac{r_0}{\alpha}\left(1-e^{-\alpha{\mathcal{V}}/r_{0}}\right), \qquad r\eqdef Re^{-\alpha{\mathcal{V}}/r_{0}}\label{TR-transf}\ ,
		\end{align}
with the explicit form of the metric in $(\mathcal{V},R)$ coordinates given in Appendix \ref{coord-transf}.
		Then, defining a time coordinate $\tilde{t}$ by
		\begin{align}
			d\tilde{t}=d\mathcal{V}- {b(R)}\left(1-\frac{r_0}{R}+\frac{2\alpha R}{r_0}\right)^{-1}dR\ , \label{dtTR}
		\end{align}
allows the metric to take the form
\be
ds^2=\sg_{\tilde t\tilde t}d\tilde{t}^2+2\sg_{\tilde t r} d\tilde{t}dr+\sg_{rr}dr^2+r^2d\Omega_{2}\ ,
\ee
where expressions for the metric components  $\sg_{\mu\nu}(\tilde{t},r)$ are again given in Appendix \ref{coord-transf}. The coordinates $\cal{V}$ and $R$ that appear therein are treated as functions of $\tilde t$ and $r$. The function $b(R)$ is then chosen such that the off-diagonal metric component $\sg_{\tilde t r}$ vanishes:
\be
	  b(R) =  \left(1-\frac{r_0}{R}+ \frac{2\alpha R}{r_0}\right)\left(1-\frac{r_0}{R}+ \frac{\alpha R}{r_0}\right)^{-1} \label{bR}
\ee
As a result, the metric becomes
\begin{align}\label{met2}
	ds^2=-e^{-2\alpha \mathcal{V}/r_0} \frac{\left(1-\frac{r_0}{R}+ \frac{\alpha R}{r_0}\right)^2}{1-\frac{r_0}{R}} d\tilde{t}^2
	+ \frac{dr^2}{1-\frac{r_0}{R}}+ r^2 d\Omega_2\ .
\end{align}
Comparing \eqref{met2} with the general spherically symmetric metric \eqref{m:tr} identifies the metric function $f$ as
\begin{align}
	f(\tilde{t},r)=1-\frac{C(\tilde{t},r)}{r}=1-\frac{r_{+}(v)}{r}=1-\frac{r_0}{R}\ ,
\end{align}
with $h$ being given by
\begin{equation}
e^{\tilde{h}(\tilde{t},r)}= e^{-a{\mathcal{V}}/r_0} \left(\frac{1-\frac{r_0}{R}+ \frac{\alpha R}{r_0}}{1-\frac{r_0}{R}}\right)\ .
\end{equation}

From Eqs.~\eqref{TR-transf}, \eqref{dtTR}, and \eqref{bR}, supplemented by the initial condition  $\tilde{t}\big(r=r_{0},v=0\big)=0$,  we obtain
\begin{align}\label{tP}
	\tilde t(v,r)&=\frac{r_0}{2\alpha}\ln\left(\frac{\alpha r_0^2}{\alpha r^2-r_+^2+r_+ r}\right) \\
&\quad+\frac{r_0}{\alpha\sqrt{1+4\alpha}}\,\mathrm{arctanh}\left(\frac{\sqrt{1+4\alpha}(r-r_+)}{(1+2\alpha)r-r_+}\right)\ .\nonumber
	\end{align}
Note that it is still possible to apply an arbitrary coordinate transformation $\tilde t\to t=T(\tilde t)$. The choice can be constrained by considering the form of the relations between $v$, $t$ and $r$ in the asymptotic region.

		Using Eq.~\eqref{tP}, the limit of $\tilde{t}$ as it propagates backwards along an ingoing null geodesic (i.e $v$ is constant and $ r\to \infty$) is
		\begin{align}
			\tilde t&\to \frac{r_0}{\alpha}\left[-\ln\left(\frac{r}{r_0}\right)+\gamma\right]\ , \label{tit}
		\end{align}
	where we have defined
	\begin{align}
		\gamma:=\frac{1}{\sqrt{1+4\alpha}}\,\mathrm{arctanh}\left(\frac{\sqrt{1+4\alpha}}{1+2\alpha}\right)\ .
	\end{align}
		Similarly we find that
		\be
		\tilde{h}\to \ln\left(\frac{\alpha r}{r_0}\right)\ .
		\ee
Since we require an asymptotic relation $v\approx t+r$ and
		\be
		r\to r_0\exp\left(-\frac{\alpha\tilde{t}}{r_0}+\gamma\right)
		\ee
		for $\tilde t\to -\infty$, we define the new time variable as
		\be
		t\defeq - r_0\exp\left(-\frac{\alpha\tilde{ t}}{r_0}+\gamma\right)+\mathfrak{t}\ .
		\ee
We choose the constant $\mathfrak{t}$ so that $t=0$ at $\tilde{t}=0$, hence
		\be
		\mathfrak{t}=r_0e^{\gamma}\ .
		\ee
		Noting that
		\be
		T'(\tilde{t})=-\frac{\alpha}{r_0}(t-\mathfrak{t})\ ,
		\ee
		we see that $h(t,r)\to 0$ at constant $v$ and $r\to \infty$, while
		\be
		\frac{dr_\sg}{dt}=\frac{r_\sg}{t-\mathfrak{t}}\ ,
		\ee
		resulting in the linear evaporation law
		\be
		r_\sg(t)=r_0-e^{-\gamma} t\ .
		\ee
		We have thus presented an exactly solvable model for an evaporating PBH, using  a conformal transformation based on a linear evaporation law.
		
\section{Physical black holes in cosmology}\label{cosmo}

Models of compact objects with cosmological boundary conditions have been investigated since the introduction of
the  McVittie metric \cite{Mv:933}, which generalizes the Schwarzschild solution to
arbitrary FRLW spacetimes.  In isotropic coordinates $(t,\bar r)$ the McVitte metric has the form
\begin{align}
ds^2=&-\frac{\left(1-m(t)/2\bar r\right)^2}{ \left(1+m(t)/2\bar r\right)^2}dt^2\nonumber \\
&+a^2(t)\left(1+\frac{m(t)}{2\bar r}\right)^4\left(d\bar r^2+\bar r^2d\Omega_2\right),
\end{align}
where the time dependence of the mass function is governed by the scale factor $a(t)$ such that
\be
m(t)\equiv m_0/a(t)\ . \label{MVsca}
\ee

Due to the non-linearity of the Einstein equations, it is impossible to split a metric into a homogenous and isotropic cosmological   background  and a part describing even a spherical inhomogeneity.  It is possible, however, to loosely describe the ``embedding” of black holes into a cosmological ``background” if the metric reduces to a FRLW metric when the parameter that describes the inhomogeneity vanishes \cite{vF:15}.  Despite the existence of numerous models with quite remarkable properties \cite{vF:15,FGF:21}, there currently exist no MBH solutions that satisfy observational constraints at the horizon scale while approaching an FRLW metric on the largest scales \cite{FCZT:23}.

We begin the procedure of embedding a PBH into a spatially flat FRLW background by writing its metric in the form of Eq.~\eqref{m:tr} \cite{vF:15}. In comoving coordinates $(\bar t,\chi)$ one has
\begin{equation}
ds^2= -d\bar t^2+ a^2(\bar t) \left(d\chi^2+ \chi^2 d\Omega_2\right)\ ,
\end{equation}
while using the areal radius $r$ as the radial coordinate brings the metric into Painlev\'e-Gullstrand form
\begin{equation}
ds^2= -\left(1- H^2 r^2\right) d\bar t^2- 2 H r d\bar t dr+ dr^2+ r^2 d\Omega_2\ ,
\end{equation}
where $H=\dot a/a$ is the Hubble parameter.
The cross term can be eliminated and the metric can be written in Schwarzschild form by introducing a new time coordinate $t$ for which
\begin{equation}
dt= \frac{1}{F} \left(d\bar t+ \beta dr\right)\ , \label{cv106}
\end{equation}
where $F(t,r)$ is an integration factor satisfying
\begin{equation}
\frac{\partial}{\partial r} \left(\frac{1}{F}\right)= \frac{\partial}{\partial \bar t} \left(\frac{\beta}{F}\right), \label{if107}
\end{equation}
and the function  $\beta$ is chosen so that $\sg_{tr}=0$. This is accomplished by having
\begin{equation}
\beta= \frac{H r}{1- H^2 r^2}\ ,
\end{equation}
which results in the line element
\begin{equation}
ds^2= -\left(1- H^2 r^2\right) F^2 dt^2+ \frac{1}{1- H^2 r^2} dr^2+ r^2 d\Omega_2\ . \label{le108}
\end{equation}
This is the spatially flat FLRW metric in Schwarzschild coordinates. De Sitter space is the special case where $H\equiv\text{const.}$ and $F\equiv 1$.

The above result easily follows from Eqs.~\eqref{eq:Gtt}--\eqref{eq:Grr},  where we set
\be
T_{\mu\nu}=T_{\mu\nu}^\mathrm{mat}-\Lambda \sg_{\mu\nu}/8\pi\ , \label{eq:Gtt-L}
\ee
separating the EMT into the matter and the cosmological vacuum parts, respectively. Then Eqs.~\eqref{eq:Gtr} and \eqref{eq:Grr} remain unchanged, apart from $\tau_a\to \tau_a^\mathrm{mat}$, while Eq.~\eqref{eq:Gtt} takes the form
\be
\pad_r C =8\pi \tau^{\mathrm{mat}}_t/f+\Lambda r^2\ .
\ee
Setting $\tau_a=0$ then  results in Eq.~\eqref{le108} with $F=1$ and $H=\sqrt{\Lambda/3}$. The cosmological constant does not affect the structure of the function $C(t,r)$, but the Schwarzschild radius is modified as
\be
r_\sg\to r_\sg(1+H^2 r_\sg^2/3)+\cO(H^4)\ ,
\ee
 while the expansions of the metric functions $C$ and $h$ retain the same form as before.

In the case of a spatially flat de Sitter space, the  advanced and retarded null coordinates become generalizations of the Eddington--Finkelsein coordinates.  They can be defined analogously to the Schwarzschild spacetime. For example, using the advanced null coordinate
\be
v\defeq t+r_*\ , \qquad r_*\defeq \frac{1}{2H}\frac{1+r/H}{1-r/H}\ ,
\ee
where $r_*$ is the de Sitter analogue of the tortoise coordinate, the de Sitter metric  can be written as Eq.~\eqref{m:vr} with
\be
 h_+(v,r)=0\ ,\qquad f(v,r)=1-H^2 r^2\ .
\ee

It is natural to consider Vaidya black holes, both using retarded \cite{M:85} and advanced \cite{M:86} null coordinates. The generalization is most easily obtained from the Einstein equations in $(v,r)$ or $(u,r)$  (see Appendices \ref{appendix-vr} and \ref{appendix-ur}), where the only nonvanishing component of the matter EMT is the standard Vaidya term  $T^\mathrm{mat}_{vv}=m'/(4\pi r^2)$. In the $(v,r)$ case this results in
\begin{align}
	f(v,r)=1-\frac{2m(v)}{r}- H^2 r^2\ , \label{VM}
\end{align}
and $h_+\equiv0$ with $\Lambda=3H^2$.

In a cosmological setting, $m\ll H^{-1}$ and the Schwarzschild radius is slightly modified by the cosmological coupling. Treating the model of Eq.~\eqref{VM} as a PBH we find
\be
r_+(v)= 2m \big(1+4m^2 H^2+\cO(H^4)\big)\ ,
\ee
similar to the Schwarzschild--de Sitter metric \cite{vF:15}.  It is also interesting to note that
\be
w_1=3r_+^2H^2=12m^2 H^2+\cO(H^4)\ ,
\ee
which shows a deviation from a Vaidya-like geometry due to the presence of the cosmological horizon. This also disagrees with the static Schwarzschild limit, consistent with the modifications to the ordinary first law which are required when considering asymptotically de Sitter (or anti-de Sitter) black holes \cite{kastor09,wald92}.

We now consider the embedding of physical black holes in general cosmological spacetimes, which is most conveniently described in Schwarzschild coordinates.  The presence of a trapped region does not impose any additional conditions, while on approach to the cosmological apparent horizon the metric takes the form of Eq.~\eqref{le108},
\be
e^h\to F\ , \qquad C\to H^2 r^3\ ,
\ee
so in addition to the outer apparent horizon which bounds the trapped region at $r=r_\sg(t)$, there is a cosmological apparent horizon at
\be
r\approx \frac{1}{H}\ .
\ee
As a result, all the properties of the near-horizon geometry described above remain valid. In particular, the expansions of the metric functions remain the same, and a region where the null energy condition is violated is expected to form near the outer apparent horizon. A schematic Carter-Penrose diagram for a RBH in an asymptotically flat de Sitter spacetime is shown in   Fig.~{\ref{rbh2}}(b).

\section{Discussion}\label{discus}

Our results show that for an uncharged physical black hole, compliance with the first law of black hole mechanics results in the coincidence of the Hayward--Kodama surface gravity with its Schwarzschild black hole value $\kappa_\mathrm{K}=1/(2r_\sg)=1/4M$, and implies that the metric near the outer apparent horizon is approximately Vaidya (that $w_1=0$). For a charged black hole it is  possible to match the surface gravity with that of the Reissner--Nordstr\"{o}m black hole,
\be
\kappa=\frac{r_+-r_-}{2r_+^2}\ ,
\ee
where $r_+$ and $r_-$ are the areal radii of the outer and the inner horizons, respectively, by having $w_1\neq 0$ \cite{MS:23}.
While it is obvious that Page's law can be maintained in both $(t,r)$ and $(v,r)$ coordinate systems in identical form, it is unclear if this is compatible with the redefinition of the null coordinate $v$ required to have $\zeta_0=0$.

We also have seen that a simple and pathology-free PBH model in asymptotically de Sitter space does not adhere to the ordinary form of the first law. This can be seen as a natural consequence of the inclusion of back-reaction in our model. In asymptotically de Sitter black hole spacetimes, it is known that a first law can be formulated separately for the event and cosmological horizons \cite{hawking77}. However, if back-reaction from the Hawking flux of each horizon is not ignored, the heat flux between the two horizons places the system out of equilibrium and the first law no longer suffices to capture variations between nearby equilibrium states for the entire spacetime. This back-reaction issue (along with ambiguities in the definition of mass in de Sitter spacetimes \cite{dolan13}) makes formulating the laws of black hole mechanics in de Sitter technically and conceptually challenging, though a number of solutions have been extensively pursued \cite{urano09,sim16,sim19}. While it may be superfluous, we stress that thermodynamic considerations for non-equilibrium systems without backing from microscopic calculations or a non-equilibrium framework should be treated with care.

Modelling black holes as PBHs results in a number of important peculiarities. The NEC violation (and existence of a macroscopic domain with negative energy density) is a necessary consequence of the formation of a trapped region in finite time according to a distant observer. This property is shared with many ECO models. However, spherically-symmetric PBHs that purport to model zero angular momentum ABHs  do not allow growth; only a solution with decreasing $r_\sg(t)$ is possible. Thus we conclude with the following chain of conditional statements: spherically-symmetric ABHs do not grow, or if they do then either they are horizonless objects or semiclassical gravity breaks down at the horizon scale.

It remains to be understood how the near-horizon EMT is compatible with current cosmological observations, such as those reported in Ref.~\cite{FCZT:23}, and if/how this EMT can be generated by relatively weak (at macroscopic scales) quantum effects \cite{FN:98,M:15}. Moreover, it is unclear how the presence of regions with negative energy density and/or pressure is compatible with absorption of the cosmic microwave background radiation.

While dealing with axially-symmetric PBHs is much more difficult \cite{DT:20}, the investigation of their embedding in a cosmological background is very important. In forthcoming work, we will detail the embedding of Kerr--Vaidya metrics in asymptotically de Sitter spacetimes. These models will serve as a basis for developing even more sophisticated descriptions of dynamical physical black holes, and provide a framework for extracting observational features of their astrophysical manifestations. These issues will be addressed in part II and III of this series.

		\acknowledgments
		Useful discussions with Robert Mann, Swayamsiddha Maharana, Sebastian Murk and Amos Ori are gratefully acknowledged. P.K.D. and I.S. are supported by an International Macquarie
		University Research Excellence Scholarship. The work of
		D.R.T. is supported by the ARC Discovery project Grant No.
		DP210101279. F.S. is funded by the ARC Discovery project Grant No. DP210101279.
		
		\appendix
\appendix

		\section{SUMMARY OF USEFUL RELATIONS}
		\subsection{Effective EMT components in $(t,r)$ coordinates}

		We give a detailed summary of the relations used in this paper in $(t,r)$ coordinates. By explicitly including higher-order terms in the expansions of the MSH mass $C(t,r)$ and metric function $h(t,r)$, the Einstein equations \eqref{eq:Gtt}, \eqref{eq:Gtr}, and \eqref{eq:Grr} give the form of various EMT components to comparative order. The expansion of $C(t,r)$ is given by
		\begin{align}
		C(t,r)=r_{g}(t)+c_{12}(t)\sqrt{x}+c_{1}(t)x+\cO(x^{3/2})\ ,
		\end{align}
		with $x=r-r_{g}(t)$ and coefficients given by
		\begin{align}
		c_{12}(t)=-4\sqrt{\smash{\pi r_\sg^3}\mathstrut}\Upsilon\ , \quad c_{1}(t)=\frac{1}{3}+\frac{4\sqrt{\smash{\pi r_\sg^3}\mathstrut}e_{12}}{3\Upsilon}\ .
		\end{align}
		The expansion of $h(t,r)$ is likewise given by
		\begin{align}
		h(t,r)=-\frac{1}{2}\ln{\frac{x}{\xi(t)}}+h_{12}(t)\sqrt{x}+\cO(x)\ ,
		\end{align}
		where
		\begin{align}
		h_{12}(t)=\frac{1}{2\sqrt{\pi}r^{3/2}_{g}\Upsilon}-\frac{e_{12}-3p_{12}}{6\Upsilon^2}\ .
		\end{align}
		The effective EMT components defined in Section \ref{pbh} then have the following series expansions
		\begin{align}
		\tau_{t}&=-\Upsilon^2+e_{12}(t)\sqrt{x}+e_{1}(t)x+\cO(x^{3/2})\ ,\label{tau-t}\\
		\tau^{r}_{t}&=-\Upsilon^2+\phi_{12}(t)\sqrt{x}+\phi_{1}(t)x+\cO(x^{3/2})\ ,\label{tau-tr}\\
		\tau^{r}&=-\Upsilon^2+p_{12}(t)\sqrt{x}+p_{1}(t)x+\cO(x^{3/2})\ ,\label{tau-r}
		\end{align}
		where
		\begin{align}
		\phi_{12}=\frac{1}{2}(e_{12}+p_{12})\ .
		\end{align}

		\subsection{Einstein equations and effective EMT components in $(v,r)$ coordinates}\label{appendix-vr}

		In $(v,r)$ coordinates, the EMT is represented as $\Theta_{\mu\nu}$ and the effective EMT components are
		\begin{align}
		\theta_{v}=e^{-2h_{+}}\Theta_{vv}\ ,\quad \theta_{vr}=e^{-h_{+}}\Theta_{vr}\ ,\quad \theta_{r}=\Theta_{rr}\ .
		\end{align}
		The Einstein equations then take the following form
		\begin{align}
		\partial_{v}C_{+}&=8\pi r^2e^{h_{+}}(\theta_{v}+f\theta_{vr})\ ,\label{dvCp}   \\
		\partial_{r}C_{+}&=-8\pi r^2\theta_{vr} \label{drCp}\ ,\\
		\partial_{r}h_{+}&=4\pi r\theta_{r}\ .
		\end{align}

		Using  the coordinate transformation \eqref{trvr-transformation} one can find relations between the effective EMT components in $(v,r)$ with those in $(t,r)$. They are related through
		\begin{align}
		\theta_{v}=\tau_{t},\quad \theta_{vr}=\dfrac{\tau^{r}_{t}-\tau_{t}}{f}\ ,\quad \theta_{r}=\dfrac{\tau_{t}+\tau^{r}-2\tau^{r}_{t}}{f^{2}}\ .\label{eff-EMT-tr-vr}
		\end{align}
		Expanding the LHS of Eq. \eqref{drCp} in a series around $r_{+}$ and the RHS around $r_{g}$, after making use of Eq. \eqref{eff-EMT-tr-vr}, and comparing order-by-order, one arrives at the following relation for $w_{1}(v)$:
		\begin{align}
		w_{1}(v)=\frac{e_{12}-p_{12}}{\Upsilon}\sqrt{\pi}r^{3/2}_{g}\label{w1-t}
		\end{align}
		The condition $e_{12}(t)=p_{12}(t)$ is therefore equivalent to $w_{1}(v)=0$. From Eq. \eqref{dvCp} in the near horizon limit, we get a relation for the evaporation rate
		\begin{align}
		e^{-\zeta_0}r'_{+}(v)=\frac{8\pi r^{2}_{+}\theta^{+}_{v}}{1-w_{1}}\ ,
		\end{align}
		where $\theta^{+}_{v}:=\lim_{r\to r_\sg}\theta_{v}=-\Upsilon^{2}$. In the final equality we used relation \eqref{eff-EMT-tr-vr}. With appropriate redefinition of the advanced coordinate $v$ one can eliminate the exponential term $e^{\zeta_0}$ and arrive at
		\begin{align}
		r'_{+}(v)=-\frac{8\pi r^{2}_{+}\Upsilon^2}{1-w_{1}}\label{rpp}\ ,
		\end{align}
		where we have used the same variable $v$ for the redefined coordinate.
		
		\subsection{Details of the coordinate transformation}\label{coord-transf}
		The linearly evaporating Vaidya metric is given by Eq. \eqref{Vaidya}. We will perform a coordinate transformation from $(v,r)$ to $(\mathcal{V},R)$ coordinates, where these coordinates are defined in Eq.\eqref{TR-transf}. The transformed metric is given by
		\begin{align}
		ds^2=e^{-2\alpha \mathcal{V}/r_{0}}\left(-\left(1-\frac{r_{0}}{R}+\frac{2\alpha R}{r_0}\right)d\mathcal{V}^2+2d\mathcal{V}dR+R^2d\Omega_{2}\right).
		\end{align}
		We then define a timelike coordinate $\tilde{t}$ by
		\begin{align}
		d\tilde{t}=d\mathcal{V}- {b(R)}\left(1-\frac{r_0}{R}+\frac{2\alpha R}{r_0}\right)^{-1}dR , 
		\end{align}
		that allows one to re-write the metric as
		\be
		ds^2=\sg_{\tilde t\tilde t}d\tilde{t}^2+2\sg_{\tilde t r} d\tilde{t}dr+\sg_{rr}dr^2+r^2d\Omega_{(2)}\ ,
		\ee
		where explicit values for the metric components  $\sg_{\mu\nu}(\tilde{t},r)$ are given below, and the coordinates $\cal{V}$ and $R$ that appear are treated as functions of $\tilde t$ and $r$. The function $b(R)$ is chosen by requiring that the off-diagonal metric component $\sg_{\tilde t r}$ vanishes. Furthermore, by using \eqref{TR-transf} and \eqref{dtTR} the differential $dR$ can be written as follows
		\begin{align}
		dR=\frac{A(r)}{1-\frac{r_{0}}{r}+\frac{(2-b(R))\alpha R}{r_0}}\left(e^{\alpha{\mathcal{V}}/r_{0}}dr+\frac{\alpha R}{r_0}d\tilde{t}\right)\ ,
		\end{align}
		where for simplicity we have defined
		\begin{align}
		A(r)=1-\frac{r_0}{R}+\frac{2\alpha R}{r_0}\ .
		\end{align}
		The metric becomes
		\begin{widetext}
			\begin{align}
			&ds^2= e^{-\tfrac{2\alpha \mathcal{V}}{r_0}}\! \left[\Bigg(-A(r)+ \frac{ 2(1- b(R))A(r)}{1-\frac{r_0}{R}+ \frac{(2-b(R))\alpha R}{r_0}} \frac{\alpha R}{r_0}
			+\frac{(2 b(R)- b(R)^2)A(r)}{\left(1-\frac{r_0}{R}+ \frac{(2-b(R))\alpha R}{r_0}\right)^2} \frac{a^2 R^2}{r_0^2}\Bigg) d\tilde{t}^2+\right.\\
			&\left.+ 2 e^{\tfrac{a \mathcal{V}}{r_0}}\! \left(\frac{(1- b(R))A(r)}{1-\frac{r_0}{R}+ \frac{(2-b(R))a R}{r_0}}+  \frac{(2 b(R)- b(R)^2)A(r)}{\left(1-\frac{r_0}{R}+ \frac{(2-b(R))\alpha R}{r_0}\right)^2} \frac{a R}{r_0}\right)d\tilde{t} dr+
			e^{2\alpha \mathcal{V}/r_0} \frac{(2 b(R)- b(R)^2)A(r)}{\left(1-\frac{r_0}{R}+ \frac{(2-b(R))\alpha R}{r_0}\right)^2} dr^2+ R^2 d\Omega_2\right].\nonumber	
			\end{align}
		\end{widetext}
		Requiring the coefficient of the $d\tilde{t} dr$ term to vanish gives the form of the function $b(R)$ as in Eq.\eqref{bR}. As a result, the metric simplifies to Eq.\eqref{met2} and the evaporation rate becomes 
		\begin{align}
		\frac{dr_{g}}{d\tilde{t}}=-\frac{\alpha r_{g}}{r_0}\ ,
		\end{align}
		while from the relation \eqref{Ups-Vaidya}, assuming a linear evaporation law $\Gamma_{+}=\alpha$, we have that
		\begin{align}
		\Upsilon=\frac{\sqrt{\alpha}}{2\sqrt{2\pi} r_{g}}\ .
		\end{align}
		Additionally, using the relation \eqref{eq:k0rp} one can show that
		\begin{align}
		\tilde{\xi}(\tilde{t})=\frac{\alpha^2r_\sg}{16\pi r^2_{0}\Upsilon^2}\ .
		\end{align}

		As a consistency check, we can rewrite the expression for $\tilde{h}$ in the vicinity of the apparent horizon,
		\be
		\tilde{h}(\tilde{t},r)\approx\ln\frac{\alpha r_+^2}{r_0(r-r_+)}=\ln\frac{\alpha C^2 }{r_0 r f}\to \ln\frac{\alpha\sqrt{r_\sg}}{4\sqrt{\pi}r_0\Upsilon\sqrt{x}},
		\ee
		and confirm the validity of the expression for $\bar\xi$. A direct evaluation gives that Eq.~\eqref{k0h} holds identically.

		\section{USEFUL RELATIONS IN RETARDED COORDINATES}\label{ur-section}
		\subsection{Series expansion of the metric functions in $(u,r)$ coordinates}\label{appendix-ur}

		The line element of the metric in $(u,r)$ coordinates is given by
		\begin{align}
		ds^2=-e^{2h_{-}(u,r)}f(u,r)du^2-e^{h_{-}(u,r)}dudr+r^2d\Omega_2\ ,
		\end{align}
		where
		\begin{align}
		f(u,r)=1-\frac{C_{-}(u,r)}{r}\ ,
		\end{align}
		with $C_{-}(u,r)$ representing the invariant MSH mass. The transformation laws from $(t,r)$ to $(u,r)$ coordinates are given by
		\begin{align}
		dt=e^{-h(t,r)}\left(e^{h_{-}(u,r)}du^2+\frac{dr}{f}\right)\ .\label{transformation-trur}
		\end{align}
		The transformation law between $(v,r)$ and $(u,r)$ coordinates is obtained by combining equations \eqref{trvr-transformation} and \eqref{transformation-trur}, giving
		\begin{align}
		du=e^{-h_{-}(u,r)}\left(e^{h_{+}(v,r)}dv-\frac{2}{f}\right)\ .\label{transformation-vrur}
		\end{align}

		The Einstein equations in $(u,r)$ coordinates are
		\begin{align}
		&-e^{-h_{-}}\partial_{u}C_{-}+f\partial_{r}C_{-}=8\pi r^2\bar{\theta}_{u}\label{duCm}\ ,\\
	     &\partial_{r}C_{-}=8\pi r^2\bar{\theta}_{ur}\label{drCm}\ ,\\
		&\partial_{r}h_{-}=4\pi r\bar{\theta}_{r}\label{drhm}\ ,
		\end{align}
		where the effective EMT components are defined as
		\begin{align}
		\bar{\theta}_{u}=e^{-2h_{-}}\bar{\Theta}_{uu}\ , \quad \bar{\theta}_{ur}=e^{-h_{-}}\bar{\Theta}_{ur}\ , \quad \bar{\theta}_{r}=\bar{\Theta}_{rr}\ .\label{theta-bar}
		\end{align}
		Before proceeding with solving Einstein's equations, it is useful to write down the equations relating the effective EMT components in $(u,r)$ with the other coordinate systems. This can be done by transformation of the EMT components according to the laws \eqref{transformation-trur} and \eqref{transformation-vrur}. The relations are
		\begin{align}
		\bar{\theta}_{u}=\tau_{t}\ ,\quad \bar{\theta}_{ur}=\frac{\tau_{t}+\tau^{r}_{t}}{f}\ ,\quad  \bar{\theta}_{r}=\frac{\tau_{t}+\tau^{r}+2\tau^{r}_{t}}{f^{2}}\ ,\label{theta-tau}
		\end{align}
		and
		\begin{align}
		\bar{\theta}_{u}=\theta_{v}\ , \quad \bar{\theta}_{ur}=\theta_{vr}+\frac{2\theta_{v}}{f}\ ,\quad \bar{\theta}_{r}=\frac{4\theta_v+4f\theta_{vr}+f^2\theta_{r}}{f^{2}}\ .
		\end{align}
		Solving the Einstein equations begins with the use of Eq. \eqref{drCm}. We seek a solution of the form
		\begin{align}
		C_{-}(u,r)=r_{-}(u)+W_{-}(u,r)\ ,
		\end{align}
		with $W_{-}(u,r_{-})=0$, so that $r_{-}$ represents the apparent horizon radius in $(u,r)$ coordinates. We also define the variable $z:=r-r_{-}$, so that near the apparent horizon we have the following partial differential equation
		\begin{align}
		\partial_{z}W_{-}=\frac{-16\pi r^{3}_{-}\Upsilon^2}{z-W_{-}}\ ,
		\end{align}
		which admits the following series solution
		\begin{align}
		W_{-}(u,r)=-4\Upsilon\sqrt{2\pi r^{3}_{-} }\sqrt{z}+\cO(z)\ .
		\end{align}
		The expansion of the MSH mass will be written in a similar form as $C(t,r)$,
		\begin{align}
		C_{-}(u,r)=r_{-}(u)+\bar{c}_{12}(u)\sqrt{z}+\bar{c}_{1}(u)z+\cO(z^{3/2})\ ,\label{cm}
		\end{align}
		with
		\begin{align}
		\bar{c}_{12}(u)=-4\Upsilon\sqrt{2\pi r^{3}_{-} }\ .
		\end{align}
		We continue with the solution of Eq. \eqref{drhm}, which near the horizon (in terms of the variable $z$) becomes
		\begin{align}
		\partial_{z}h_{-}=-\frac{1}{2z}+\cO\left(\frac{1}{\sqrt{z}}\right)\ .
		\end{align}
		We thus obtain the solution
		\begin{align}
		h_{-}(z,r)=-\frac{1}{2}\ln{\frac{z}{\bar{\xi}(u)}}+\bar{h}_{12}(u)\sqrt{z}+\bar{h}_{1}(u)z+\cO(z^{3/2})\ .\label{hm}
		\end{align}
		The functions $h(t,r)$ and $h_{-}(u,r)$ exhibit the same logarithmically divergent behaviour at the horizon, leading to the aforementioned issues with $(t,r)$ and $(u,r)$ coordinates. We now proceed with the final Einstein equation \eqref{duCm}, which plays the role of a consistency condition since $C_{-}$ and $h_{-}$ have already been determined. This consistency condition is used to extract information about the evaporation rate in $(u,r)$ coordinates. Using the solutions for $C_{-}$ and $h_{-}$ in \eqref{duCm}, and taking the near-horizon limit, we have that
		\begin{align}
		r'_{-}(u)=-2\sqrt{2\pi r_{-}\bar{\xi}(u)}\Upsilon\ .\label{rmp}
		\end{align}

		\subsection{Coordinate transformations}
		To derive the transformation between $(t,r)$ and $(u,r)$ coordinates, we proceed in a fashion similar to the $(v,r)$ case, expressing the variable $z=r-r_-$ in terms of $y$ and $x$. The computations should be done along an ingoing null geodesic due to the non-singular behaviour of the advanced coordinate $v$. We start by first considering the relation between $z$ and $y$, which requires determining the variation of $u$ along the ingoing null geodesic. $u$ can be written as a function of $v$ and $r$ using the transformation \eqref{transformation-vrur},
		\begin{align}
		u(v,r)=u(v,r_{+})+(\partial_{r}u)\!\!\!\underset{y=0}{\big|}y+\tfrac{1}{2}(\partial^2_{r}u)\!\!\!\underset{y=0}{\big|}y^{2}+\cO(y^3),
		\end{align}
		or in a simpler form
		\begin{align}
		\delta u=(\partial_{r}u)\!\!\!\underset{y=0}{\big|}y+\tfrac{1}{2}(\partial^2_{r}u)\!\!\!\underset{y=0}{\big|}y^{2}+\cO(y^3)\ .
		\end{align}
		From the transformation law \eqref{transformation-vrur} and the expansions \eqref{cm} and \eqref{hm} we have that
		\begin{align}
		\partial_{r}u\!\!\!\underset{y=0}{\big|}=\left(-e^{-h_{-}}f^{-1}\right)\!\!\!\underset{y=0}{\big|}=\frac{1}{r'_{-}(u)}\ .
		\end{align}
		The variation of $u$ can thus be written as
		\begin{align}
		\delta u=\frac{y}{r'_{-}}+\tfrac{1}{2}(\partial^2_{r}u)\!\!\!\underset{y=0}{\big|}y^{2}+\cO(y^3)\ .\label{delta u}
		\end{align}
		Now we can proceed with the calculation of the relation between $z$ and $y$. We define $z$ as a function of $v$ and $r$ as
		\begin{align}
		z(v,r_{+}+y)=(r_{+}+y)-r_{-}(u(v,r_{+}+y))\ .\label{zy-gen}
		\end{align}
		The term $r_{-}(u(v,r_{+}+y))$ is expanded as
		\be\begin{aligned}
			r_{-}(u(v,r_{+}+y))&=r_{-}(u(v,r_{+}))+r'_{-}(u)\delta u\\
			&\quad+\tfrac{1}{2}r''_{-}(u)\delta u^2\ .
		\end{aligned}\ee
		Identifying $r_{-}(u(v,r_{+}))=r_{+}(v)$ and using the above equation in \eqref{zy-gen} we find that near the apparent horizon $z$ and $y$ are related through
		\begin{align}
		z=\frac{1}{2}\tilde{\omega}^2y^2\ ,\label{zy}
		\end{align}
		with
		\begin{align}
		\tilde{\omega}^2=-r'_{-}(u)(\partial^2_{r}u)\!\!\!\underset{y=0}{\big|}-\frac{r''_{-}(u)}{(r'_{-}(u))^2}\ .\label{omu}
		\end{align}
		The derivative $(\partial^2_{r}u)|_{y=0}$ is finite and its determination is given in Appendix \ref{appendix-d2udr2}. To determine the relationship between $x$ and $z$ we use the relations \eqref{xy} and \eqref{zy}, which gives the following linear relationship between these coordinates:
		\begin{align}
		z=\frac{\tilde{\omega}^2}{\omega^2}x\label{gen-zx}
		\end{align}
		We can find $\bar{{\omega}}$ in the same manner as for $(t,r)$ coordinates by using the invariance of the MSH mass
		\begin{align}
		C_{-}(u(v,r),r)=C_{+}(v,r)\ .
		\end{align}
		Using both expansions of the MSH mass respectively we have
		\begin{align}
		r_{-}(u(v,r))+\bar{c}_{12}(u)\sqrt{z}=r_{+}(v)+w_{1}(v)y+\ ,\label{comp-cmcp}
		\end{align}
		with subleading terms of order $+\cO(z)$ and $\cO(y^2)$. In order to compare the left and right hand side we need to first expand $r_{-}(u(v,r))$ and then use the relation \eqref{zy}. This expansion is given by
		\begin{align}
		r_{-}(u(v,r))=r_{-}(u(v,r_{+}))+r'_{-}(u)\delta u+\cO(\delta u^2)\ .
		\end{align}
		We identify $r_{-}(u(v,r_{+}))=r_{+}(v)$ and make use of the relation \eqref{delta u}, wherein Eq. \eqref{comp-cmcp} implies
		\begin{align}
		\tilde{\omega}=\frac{1-w_{1}}{4\sqrt{\pi}r^{3/2}_{-}\Upsilon}\ . \label{varpi}
		\end{align}
		Finally, using the above equation and \eqref{w1v}, \eqref{gen-zx} becomes
		\begin{align}
		x=2z\ .\label{xz}
		\end{align}

		\subsection{Condition for $w_{1}=0$}
		The effective EMT component $\bar{\theta}_{u}$ is defined by Eq. \eqref{theta-bar}. The Einstein equations imply that
		\begin{align}
		\bar{\theta}_{u}=\frac{1}{8\pi}e^{-2h_{-}}\bar{G}_{uu}\ .
		\end{align}
		Expanding the RHS of the above equation near the apparent horizon using the expansions \eqref{cm} and \eqref{hm} gives
		\begin{align}
		\bar{\theta}_{u}=-\Upsilon^2(t)+\left(\frac{\bar{c}_1(u)\Upsilon(t)}{\sqrt{2\pi r_{-}^{3/2}}}-\bar{h}_{12}(u)\Upsilon^2(t)\right)\sqrt{z}+\cO(z)\ .
		\end{align}
		However, Eqs.\eqref{theta-tau} and \eqref{tau-t}  hold identically, so we can compare the expansions and use the relation \eqref{xz} to find that
		\begin{align}
		e_{12}(t)=\frac{1}{\sqrt{2}}\left(\frac{\bar{c}_1(u)\Upsilon(t)}{\sqrt{2\pi r_{-}^{3/2}}}-\bar{h}_{12}(u)\Upsilon^2(t)\right)\label{e12}.
		\end{align}
		The same procedure using $\tau^r$ instead gives a relation for $p_{12}(t)$. Combining equations \eqref{theta-tau} implies that
		\begin{align}
		\tau^{r}=f^2\bar{\theta}_{r}+\bar{\theta}_{u}-2f\bar{\theta}_{ur}\ .
		\end{align}
		The RHS of the above equation can be expanded about the apparent horizon in the same manner as was done for $\bar{\theta}_{u}$, by using the definition of the effective EMT components \eqref{theta-bar} and the Eqs. \eqref{cm} and \eqref{hm}. The expansion for the LHS is given by \eqref{tau-r}. Using the transformation law \eqref{xz} and comparing the expansions then gives
		\begin{align}
		p_{12}(t)=\frac{1}{\sqrt{2}}\left(-\frac{\Upsilon^2(t)}{\sqrt{2\pi r^3_{-}}}+3\bar{h}_{12}(u)\Upsilon^2(t)\right)\ .\label{p12}
		\end{align}
		The condition \eqref{w1-t}, assuming $w_{1}(v)=0$, is then equivalent to the condition that $e_{12}=p_{12}$ which immediately yields
		\begin{align}
		\bar{c}_1(u)=4\sqrt{2\pi r^3_{-}} \bar{h}_{12}(u)\Upsilon(t)-1 \label{w1=0 relation}\ ,
		\end{align}
		 when $w_1=0$.

		\subsection{Evaporation relations}
		It is useful to derive relations that connect the evaporation law in $(u,r)$ coordinates with the other coordinate systems used in this paper. We begin by writing the evaporation law in $(u,r)$, assuming it has the following form:
		\begin{align}
		r'_{-}(u)=-\Gamma_{-}(r_{-})
		\end{align}
		This implies that
		\begin{align}
		\frac{r'_{-}(u)}{r'_{g}(t)}=\frac{\Gamma_{-}}{\Gamma}
		\end{align}
		 leads to a relation between $\xi(t)$ and $\bar{\xi}(u)$,
		\begin{align}
		\bar{\xi}(u)=\frac{2\,\Gamma^2_{-}}{\Gamma^2}\,\xi(t)\ .
		\end{align}
		This relation represents a constraint between $\xi(t)$ and $\bar{\xi}(u)$, which must be satisfied in order to have the same functional form of the evaporation law in both $(t,r)$ and $(u,r)$ coordinates. A relation between the evaporation rate in $(u,r)$ and $(v,r)$ can also be found, by using Eq. \eqref{rmp} and Eq. \eqref{ups-gen}, giving
		\begin{align}
		\Gamma^2_{-}=\frac{(1-w_{1})\bar{\xi}(u)}{r_{+}}\Gamma_{+}\ .
		\end{align}

		\section{SECOND DERIVATIVES OF TIME ALONG A NULL GEODESIC}	
		\subsection{Second partial derivative in $(t,r)$}\label{appendix-d2tdr2}
		For the calculation of the second derivative which appears in $\omega$, it is necessary to use an ingoing null geodesic due to the singular nature of the coordinate $t$ at the apparent horizon. For the metric \eqref{m:tr},  ingoing null rays are described by
		\begin{align}
		\frac{dt}{dr}=-e^{-h}f^{-1}\label{drt}\ .
		\end{align}
		The second partial derivative can be written as
		\begin{align}
		\frac{\partial^2 t}{\partial r^2}\bigg|_{v}=\frac{d}{dr}\left(\frac{dt}{dr}\right)\bigg|_{v}=\frac{d}{dr}\left(-e^{-h}f^{-1}\right)\ ,
		\end{align}
		where \eqref{drt} is used in the final equality. Explicit calculation of the above equation along an ingoing null geodesic leads to
		\begin{align}
		\frac{\partial^2 t}{\partial r^2}\bigg|_{v}&=e^{-h}\left(-(\partial_{t}h)e^{-h}f^{-2}+(\partial_{r}h)f^{-1}\right.\nonumber\\
		&\left.\quad-(\partial_{t}f)e^{-h}f^{-3}+f^{-2}\partial_{r}f\right)\ .
		\end{align}
       	Using now the expansion of
       	\begin{align}
       		e^{-h}f^{-1}=\frac{1}{-r'_{g}}+\sigma \sqrt{x}+\cO{(x)},
       	\end{align}
       	where
       	\begin{align}
       		\sigma=\frac{\sqrt{\pi}(e_{12}-p_{12})r^{3/2}_g-\Upsilon}{8\pi r^2_g\xi  \Upsilon^3},
       	\end{align}
       	we can write the second partial derivative in the following form
       	\begin{align}
       		\partial^2_r t=\left(\frac{\sigma(e_{12}-p_{12})\sqrt{\pi}r^{3/2}_{g}-\sigma \Upsilon}{4\sqrt{\pi}r^{3/2}_{g} \Upsilon^2}\right)-\frac{\left(\frac{r'_g}{r_\sg}+\frac{\xi'}{\xi}+\frac{2\Upsilon'}{\Upsilon}\right)}{2(r'_g)^2}+\cO{(\sqrt{x})},\label{partial-gen}
       \end{align}
       and substitution in Eq.\eqref{xy} leads to
       \begin{widetext}
       	\begin{align}
       		\omega^2=\left[-\frac{1}{2(r'_g)^2}\left(\frac{r'_g}{r_\sg}+\frac{\xi'}{\xi}+\frac{2 \Upsilon'}{ \Upsilon}\right)-\frac{r''_g}{(r'_g)^2}\right]-r'_g\left(\frac{\sigma (e_{12}-p_{12})}{4 \Upsilon^2}-\frac{\sigma}{4\sqrt{\pi}r^{3/2}_g \Upsilon}\right)+\cO{(\sqrt{x})}.
       	\end{align}
       \end{widetext}
      The term in the square brackets will vanish if we substitute the expressions for $r'_g$ given by Eq.\eqref{eq:k0rp} and $r''_g$ which is derived by differentiating with respect to $t$ the expression of $r'_g$. So we have that
      \begin{align}
      	\omega^2=-r'_g\left(\frac{\sigma (e_{12}-p_{12})}{4 \Upsilon^2}-\frac{\sigma}{4\sqrt{\pi}r^{3/2}_g \Upsilon}\right).
      \end{align}
      Substitution of $\sigma$ leads to the following simple relation
      	\begin{align}
      	\omega^2=\frac{\left(\sqrt{\pi}(-e_{12}+p_{12})r^{3/2}_{g}+\Upsilon \right)^2}{8\pi r^3_{g}\Upsilon^4}+\cO(\sqrt{x})\ .\label{om}
      \end{align}
      	We are interested in the specific case where $w_{1}=0$, a condition which was shown in Appendix~\ref{appendix-vr} to be equivalent to $e_{12}=p_{12}$. Using this condition in Eq.\eqref{om} simplifies the result as follows:
      	\begin{align}
      		\omega^2=\frac{1}{8\pi r^{3}_{g}\Upsilon^2}\ ,
      	\end{align}
      	in accordance with Eq. \eqref{varpi-xy-w1=0}.      	
		\subsection{Second partial derivative in $(u,r)$}\label{appendix-d2udr2}
		Determining the second partial derivative entering into $\tilde{\omega}$ for $(u,r)$ coordinates proceeds in the same manner as in the previous subsection. The calculation is again performed along on an ingoing null geodesic, where the first derivative is given by
		\begin{align}
		\frac{du}{dr}=-2e^{-h_{-}}f^{-1}\label{dudr}\ .
		\end{align}	
		The second partial derivative is then
		\begin{align}
		\frac{\partial^2 u}{\partial r^2}\bigg|_v=\frac{d}{dr}\left(\frac{du}{dr}\right)\bigg|_{v}=\frac{d}{dr}\left(-2e^{-h_{-}}f^{-1}\right)\ .
		\end{align}
		Evaluating this derivative along an ingoing null geodesic leads to
		\begin{align}
		\frac{\partial^2 u}{\partial r^2}\bigg|_{v}&=e^{-2h_{-}}f^{-3}\Big(-4\partial_{u}f+2f\big(e^{h_{-}}(\partial_r f +f\partial_r h)-2\partial_{u}h_{-}\big)\Big)\ .
		\end{align}
	Using now the expansion
	\begin{align}
		e^{-h_{-}}f^{-1}=\frac{-1}{2r'_{-}}+\bar{\sigma}\sqrt{z}+\cO{(z)},
	\end{align}
	where
	\begin{align}
		\sigma=\frac{-1+\bar{c}_1-4\sqrt{2\pi}\bar{h}_{12}r^{3/2}_{-} \Upsilon}{32\pi r^2_{-} \Upsilon^2\sqrt{\bar{\xi}}},
	\end{align}
	we have that
	\begin{align}
		\partial^2_{r}u=-\frac{\left(2\sigma \bar{h}_{12}(r'_{-})^2-\sigma\sqrt{\frac{2}{\pi}}\frac{(-1+\bar{c}_1)}{4 r^{3/2}_{-} \Upsilon}(r'_{-})^2+\frac{r'_{-}}{r_{-}}+\frac{\xi'}{\xi}\right)}{2(r'_{-})^2}+\cO{(\sqrt{z})}.
	\end{align}
	Substituting this in Eq.\eqref{omu} we have
	\begin{widetext}
	\begin{align}
		\tilde{\omega}^2=\left[\frac{1}{2r'_{-}}\left(\frac{r'_{-}}{r_{-}}+\frac{\xi'}{\xi}\right)-\frac{r''_{-}}{(r'_{-})^2}\right]+\frac{1}{2r'_{-}}\left(2c\bar{h}_{12}(r'_{-})^2-c\sqrt{\frac{2}{\pi}}\frac{(-1+\bar{c}_1)}{4 r^{3/2}_{-} \Upsilon}(r'_{-})^2\right).
	\end{align}
	\end{widetext}
	The term in the brackets will vanish, after substituting the expressions for $r'_{-}$ given by Eq.\eqref{rmp} and $r''_{-}$ which is derived by differentiating $r'_{-}$ with respect to $u$, so we have
	\begin{align}
		\tilde{\omega}^2=\frac{\bar{h}^2_{12}}{2}+\frac{(-1+\bar{c}_{1})^2}{64 \pi r^3_{-} \Upsilon^2}-\frac{(-1+\bar{c}_{1})\bar{h}_{12}}{4\sqrt{2\pi}r^{3/2}_{-} \Upsilon}.
	\end{align}
	Substitution of the condition for $w_{1}=0$ which is given by Eq.\eqref{w1=0 relation} leads to
	\begin{align}
		\tilde{\omega}^2=\frac{1}{16\pi r^{3}_{-} \Upsilon^2}.
	\end{align}
This expression for $\tilde{\omega}^2$ in the near--Vaidya limit is in agreement with Eq. \eqref{varpi} for $w_{1}=0$.

\end{document}